\documentclass[sigconf]{acmart}
\AtBeginDocument{%
  }

\copyrightyear{2026}
\acmYear{2026}
\setcopyright{cc} %TODO
\setcctype{by}
\acmConference[WSDM '26] {Proceedings of the Nineteenth ACM International Conference on Web Search and Data Mining}{February 22--26, 2026}{Boise, ID, USA.}
\acmBooktitle{Proceedings of the Nineteenth ACM International Conference on Web Search and Data Mining (WSDM '26), February 22--26, 2026, Boise, ID, USA}
\acmISBN{979-8-4007-2292-9/2026/02}
\acmDOI{10.1145/3773966.3777993}

\usepackage{enumitem} % 允许自定义列表
\usepackage{booktabs} % For better table lines
\usepackage{multirow} % For multirow cells
\usepackage{tabularx} % For creating tables that fit the page width
\usepackage{graphicx}
\usepackage{array}
\usepackage{balance}
\usepackage{makecell} % 引入makecell宏包

\begin{document}

\title{EpicCBR: Item-Relation-Enhanced Dual-Scenario Contrastive Learning for Cold-Start Bundle Recommendation}


\author{Yihang Li}
\authornote{Contribute equal to this work.}
\email{liyihang1005@outlook.com}
\affiliation{%
  \institution{School of Computer Science and Technology, Huazhong University of Science and Technology}
  \city{Wuhan}
  \state{HuBei}
  \country{China}
}

\author{Zhuo Liu}
\authornotemark[1]
\email{lz20220619@163.com}
\affiliation{%
  \institution{School of Computer Science and Technology, Huazhong University of Science and Technology}
  \city{Wuhan}
  \state{HuBei}
  \country{China}
}

\author{Wei Wei}
\authornote{Corresponding author.}
\email{weiw@hust.edu.cn}
\affiliation{%
  \institution{Cognitive Computing and Intelligent Information Processing (CCIIP) Laboratory, School of Computer Science and Technology, Huazhong University of Science and Technology}
  \city{Wuhan}
  \state{HuBei}
  \country{China}
}

\begin{abstract}
  Bundle recommendation aims to recommend a set of items to users for overall consumption. Existing bundle recommendation models primarily depend on observed user-bundle interactions, limiting exploration of newly-emerged bundles that are constantly created. It pose a critical representation challenge for current bundle methods, as they usually treat each bundle as an independent instance, while neglecting to fully leverage the user-item (UI) and bundle-item (BI) relations over popular items. To alleviate it, in this paper we propose a multi-view contrastive learning framework for cold-start bundle recommendation, named EpicCBR. Specifically, it precisely mine and utilize the item relations to construct user profiles, identifying users likely to engage with bundles. Additionally, a popularity-based method that characterizes the features of new bundles through historical bundle information and user preferences is proposed. To build a framework that demonstrates robustness in both cold-start and warm-start scenarios, a multi-view graph contrastive learning framework capable of integrating these diverse scenarios is introduced to ensure the model's generalization capability. Extensive experiments conducted on three popular benchmarks showed that EpicCBR outperforms state-of-the-art by a large margin (up to 387\%), sufficiently demonstrating the superiority of the proposed method in cold-start scenario. The code and dataset can be found in the GitHub repository: \url{https://github.com/alexlovecoding/EpicCBR}.
\end{abstract}





\maketitle

\section{Introduction}

\quad Recently, modern recommender systems\cite{gao2023survey} increasingly prioritize bundle recommendation\cite{liu2017modeling} to enhance engagement through synergistic item groupings, yet grapple with newly-emerged (or called cold-start) bundles 
suffering sparse interactions—critically hindering preference modeling. This foundational challenge remains acutely under-explored. Traditional approaches (e.g., collaborative filtering [10]) frequently fail to capture nuanced user preferences amidst sparse user-bundle interactions in cold-start scenarios \cite{du2023enhancing,jeon2024cold,kim2024towards,zhao2022multi}, since without precise preference disentanglement from limited collaborative  signals \cite{du2023enhancing,jeon2024cold,kim2024towards,zhao2022multi}.

Actually, there already exists much effort devoted to bundle recommendation, such as attention mechanisms\cite{rashed2022context,ma2024personalized}, matrix factorization\cite{jang2024attention,mao2024matrix}, and graph neural networks (GNNs)\cite{sun2024survey,chang2020bundle,he2020lightgcn,gao2023survey}. Recent methods, integrating  GNNs with contrastive learning \cite{wang2024relative,zou2022multi,yu2022graph} , advances warm-start bundle recommendation scenarios through leveraging historical interactions to effectively capture collaborative signals. However, existing bundle recommendations (e.g., MultiCBR~\cite{ma2024multicbr}) easily fail in cold-start settings due to missing historical user-interactions for newly-emerged bundles. Thus, the learning of user-bundle relations alone obstructs preference signal derivation for cold-start bundles, as neither interaction-proximal users nor semantically analogous bundles exist. Later, subsequent approaches are proposed for cold-start bundle recommendation, e.g., Coheat\cite{jeon2024cold}, which uses a curriculum heating mechanism, gradually shifting the learning focus from one view to the other. However, these methods learn the bundle representation isolately, rendering recommendation process time-consuming and difficult to scale. 

In this paper, we propose EpicCBR, a popularity-driven bundle recommendation approach to  identifying users likely to engage with bundles under sparse interaction, via multi-view based contrastive representation learning in dual-scenarios, i.e., cold-start and warm-start. Specifically, we leverage user-bundle co-occurrence patterns for item-relation enhancement, which facilitates the estimation of probabilistic popularity for popular items, and  in turn can help learn the user latent selection tendencies over the highly-popular items. Furthermore, to address both cold-start and warm-start scenarios, we also specifically-design a collaborative dual-scenario framework for EpicCBR, which consists of two modules, namely, cold-start module and warm-start module. The former infers latent user interests via the enhanced item-relationships, while the latter refines explicit preferences through historical user-bundle interactions. Through contrastive learning, we align dual-scenario representations of bundles and user preferences by jointly optimizing the generalization of the model for new bundles, as well as  the recommendation accuracy on the existing bundles. This approach, through the in-depth mining of item-level semantics and the organic integration of cross-scenario information, paves a potential way for cold-start bundle recommendations. We evaluate our approach using three widely used datasets, demonstrating that EpicCBR outperforms the latest state-of-the-art (SOTA) models for cold-start bundle recommendation. The main contributions of this paper are as follows:

\begin{figure*}[t!] % [t!] 表示图片尽量放在页面顶部
    \centering
        \includegraphics[width=1.0\textwidth]{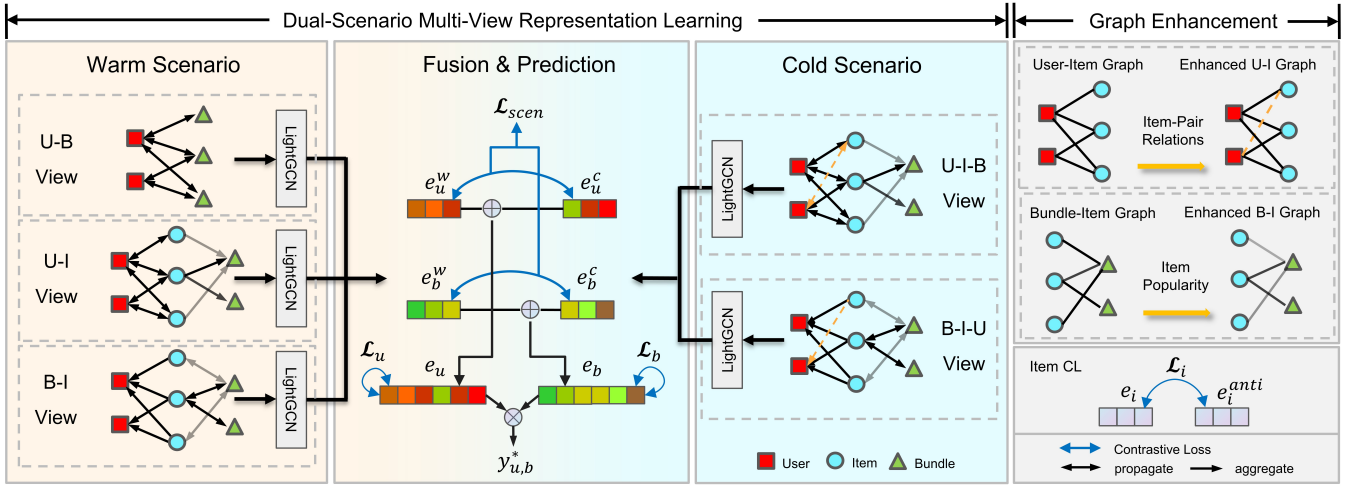} % 图片宽度设置为文本宽度
    \caption{The overall framework of EpicCBR. (1) Multi-view Representation Learning comprises a Warm-Scenario Module (User-Bundle View, User-Item View, Bundle-Item View) for historical interaction modeling and a Cold-Scenario Module (User-Item-Bundle View, Bundle-Item-User View) for deducing potential interactions. (2) Fusion \& Prediction integrates representations from both scenarios via contrastive losses to generate recommendation scores.} % 图片标题
    \label{fig:model} % 图片标签
\end{figure*}

\begin{enumerate}[label=(\alph*)]
\item To the best of our knowledge, we are the first to propose a dual-scenario framework for bundle recommendation, which integrates information from two scenario modules to ensure robust recommendations across all user-bundle interaction scenarios.
\item We propose a popularity-aware embedding generation method and an item-level view enhancement for constructing user profiles.
\item Extensive experiments on three benchmark datasets indicate that our model overperforms SOTA baselines.
\end{enumerate}

\section{Methodology}

\quad In this section, we present a bundle recommendation model designed for addressing cold-start issues, based on item-pair relations and contrastive learning. We first introduce the formal definition of the problem, and then introduce the three key modules of this model, namely, \emph{view-enhancement item-pair relation mining module} ,  \emph{popularity-aware bundle representation module}, and \emph{dual-scenario multi-view contrastive learning module}.

\subsection{Overview}
\label{sec:overview}

\quad Figure~\ref{fig:model} gives an overview of \textbf{EpicCBR}. We unify cold/warm-start bundle recommendation with two coordinated paths and a contrastive alignment:

\textbf{(1) Cold path.} We first \emph{mine item-pair relations} from the UI and BI graphs to denoise and enrich the \emph{user–item (UI) view}. In parallel, we construct a \emph{popularity-aware BI view} that assigns popularity-corrected weights to bundle–item edges for unseen bundles. Each view produces embeddings via LightGCN propagation, and the path performs view-wise early fusion.

\textbf{(2) Warm path.} We leverage historical UB/UI/BI interactions to learn user and bundle embeddings with the same LightGCN operators, ensuring architectural parity with the cold path.

\textbf{(3) Dual-scenario alignment and scoring.} We align the two paths by \emph{multi-view contrastive learning} to preserve cold-start signals without sacrificing warm-start accuracy. The final score is a unified inner-product $y^*_{u,b}=\langle e_u, e_b\rangle$ used for Top-$K_{\text{rec}}$ ranking. This design keeps heavy relation mining offline, while online inference reduces to cached-embedding retrieval and inner products.

\subsection{Problem definition}
\label{sec:problem}

\quad We study bundle recommendation over users $\mathcal{U}$, items $\mathcal{I}$, and bundles $\mathcal{B}$. Let $\mathcal{D}_{UI}\subseteq \mathcal{U}\times \mathcal{I}$ denote implicit user–item interactions, $\mathcal{D}_{BI}\subseteq \mathcal{B}\times \mathcal{I}$ denote bundle–item affiliations, and $\mathcal{D}_{UB}\subseteq \mathcal{U}\times \mathcal{B}$ denote user–bundle interactions. The objective is to learn a scoring function $y^*_{u,b}=\langle e_u,e_b\rangle$ that, under implicit feedback, ranks candidate bundles $b\in \mathcal{B}$ for a user $u$ and returns top recommendations. We focus on the \emph{cold-start} regime, where $\mathcal{B}_{\mathrm{cold}}\subseteq \mathcal{B}$ contains bundles unseen in the training user–bundle interactions, i.e., $(u,b)\notin \mathcal{D}^{\mathrm{train}}_{UB}$ for all $u\in \mathcal{U}$ and $b\in \mathcal{B}_{\mathrm{cold}}$; this places no restriction on bundle–item affiliations, meaning $\mathcal{D}^{\mathrm{train}}_{BI}$ may include $(b,i)$ with $b\in \mathcal{B}_{\mathrm{cold}}$.

\subsection{User-Item View Enhancement via Item-Pair Relation Mining}
\label{subsec:relation-mining}
\begin{table}[t]
  \caption{Item-pair relations based on UI graph and BI graph}
  \label{tab:relation}
  \begin{tabular}{l|cc}
    \toprule
    Relation& complementary(B) & exclusive(B)\\
    \midrule
    collaborative(U) & $R_1$ & $R_2$\\
    conflicting(U) & $R_3$ & $R_4$\\
  \bottomrule
\end{tabular}
\end{table}

\quad To address the challenge that UB history offers limited help in recommending new bundles, we analyze the the UI and BI graphs. For cold start bundles, we aim to pinpoint target users by examining the constituent items of the bundle. Our observations reveal that user behavior concerning item pairs encompasses complementary choices, such as as purchasing toothpaste and toothbrush simultaneously in daily life, as well as mutually exclusive choices. Furthermore, bundle strategies for item pairs also include collaborative and conflicting selections. Analysis has shown that the relations between items can be categorized into four types based on UI and BI interactions, as illustrated in Table~\ref{tab:relation}.

These relations can be effectively leveraged in bundle recommendation tasks. With our developed the item relation framework, we can create user profiles based on each user's historical interaction data. Items that share relations $R_1$ and $R_2$ with the user's interacted items are likely to be preferred by the user. Additionally, the relation $R_2$ potentially yields more information that is particularly suited for cold-start scenarios. Conversely, items that exhibit relation $R_4$ are likely unsuitable for recommendation to the user. The items in the cold start bundle can be recommended to users whose selection is highly possible, based on the constructed user profiles\cite{lu2021personalized,ma2024personalized}. Therefore, we formulate the following.

\subsubsection{Item-Pair relation Definitions}
\label{subsubsec:relation_def}
\begin{figure}[t!] % 单栏浮动体
    \centering
    \includegraphics[width=0.5\textwidth,keepaspectratio]{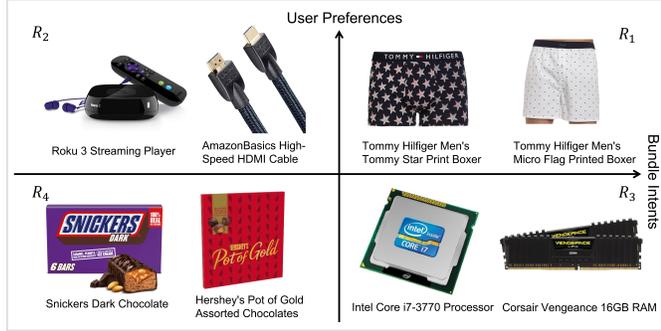} % 设置图片宽度为单栏宽度，并保持纵横比
    \caption{Visualizing Item-Pair Relations} % 图片标题
    \label{fig:item_pairs} % 图片标签
\end{figure}

An effective dual-screening strategy is employed. And we visualized the relations between items based on the definitions from three Amazon datasets: Food, Electronic, and Clothing \cite{sun2022revisiting, sun2024revisiting}. Representative item pairs from each relation are shown in Figure \ref{fig:item_pairs}, illustrating the distinct patterns between platform design and user behavior.

\textbf{Prerequisite:} For each interaction graph, the occurrence counts of the two items, $C_i$ and $C_j$, must satisfy $C_i, C_j > \theta^{I}$, where $\theta^{I}$ is a minimum frequency threshold used to remove sparse items whose co-occurrence statistics are unreliable. The Jaccard similarity is defined as:
\begin{equation}
    J_{i, j} = \frac{|E_{i, j}|}{C_i + C_j - |E_{i, j}|},
\end{equation}
where $|E_{i, j}|$ denotes the co-occurrence frequency between items $i$ and $j$ in a given interaction graph.

\textbf{Threshold Design and Motivation:}  
We adopt a dual Jaccard similarity screening with seven thresholds:
\[
\theta^{\mathrm{high}}_{UI},\ \theta^{\mathrm{low}}_{UI},\ \theta^{\mathrm{anti}}_{UI},\ 
\theta^{\mathrm{high}}_{BI},\ \theta^{\mathrm{low}}_{BI},\ \theta^{\mathrm{anti}}_{BI},\ 
\theta^{I}.
\]
Within each graph, these satisfy $\theta^{\mathrm{high}} > \theta^{\mathrm{low}} > \theta^{\mathrm{anti}}$.  
To ensure that the mined relations are both computationally efficient and highly reliable in reflecting genuine semantic associations, we set $\theta^{\mathrm{high}}$ above the 95-th percentile, $\theta^{\mathrm{low}}$ below the 20-th percentile, and $\theta^{\mathrm{anti}}$ below the 5-th percentile. Separating $\theta^{\mathrm{anti}}$ from $\theta^{\mathrm{low}}$ allows $R_4$ (anti-class) to target strongly negative or mutually exclusive pairs, rather than merely low-similarity pairs, thereby improving the discriminative power of the enhanced user–item view. These thresholds are set based on empirical analysis, which indicated this setup strikes an optimal balance between the quantity and the quality of the mined relations.

\textbf{Relation Categories:}  
Let $J^{UI}_{i,j}$ and $J^{BI}_{i,j}$ denote the Jaccard similarity between items $i$ and $j$ in the UI and BI graphs, respectively. The four relation types are defined as:
\begin{itemize}
    \item[$(1)$] \textbf{$R_1$: Same-Class Relation}  
    \emph{Definition:} $J^{UI}_{i, j} > \theta^{\mathrm{high}}_{UI}$ and $J^{BI}_{i, j} > \theta^{\mathrm{high}}_{BI}$.  
    \emph{Semantic:} Users tend to interact with both items, and the platform actively combines them, reflecting thematic or functional consistency. For example, two men’s underwear products from the same brand often recommended together.

    \item[$(2)$] \textbf{$R_2$: Cross-Domain Complementary Relation}  
    \emph{Definition:} $J^{UI}_{i, j} > \theta^{\mathrm{high}}_{UI}$ and $J^{BI}_{i, j} < \theta^{\mathrm{low}}_{BI}$.  
    \emph{Semantic:} Users spontaneously combine items across categories, while the platform does not actively design such combinations. For example, a streaming device and an HDMI cable frequently purchased together by users, but not co-bundled by the platform.

    \item[$(3)$] \textbf{$R_3$: Design-Driven Relation}  
    \emph{Definition:} $J^{BI}_{i, j} > \theta^{\mathrm{high}}_{BI}$ and $J^{UI}_{i, j} < \theta^{\mathrm{low}}_{UI}$.  
    \emph{Semantic:} The platform enforces the combination, but user acceptance is low. For example, a processor and RAM suggested together in a build package, but often purchased separately by users.

    \item[$(4)$] \textbf{$R_4$: Anti-Class Relation}  
    \emph{Definition:} $J^{UI}_{i, j} < \theta^{\mathrm{anti}}_{UI}$ and $J^{BI}_{i, j} < \theta^{\mathrm{anti}}_{BI}$.  
    \emph{Semantic:} Items exhibit mutual exclusivity or irrelevance, such as competitive products in the same category, which platforms avoid recommending together.
\end{itemize}

Our relation classification approach offers the following advantages:

1. By using item occurrence frequency as a prerequisite, we implement a triple screening process that combines occurrence frequency and dual Jaccard similarity. This ensures high credibility for the identified relations while minimizing noise from random events.
    
2. Through multilevel filtering, the number of generated relation pairs is significantly reduced, avoiding the \(n^2\) scale. This enables precise relation mining, preserves graph sparsity during structure enhancement, and facilitates subsequent applications.

\subsubsection{View Enhancement Based on Item-Pair Relations}
Based on the mined item–pair relations, we inject $R_2$ (cross-domain complementary) edges into the UI view as auxiliary structure. Concretely, we (i) build the basic heterogeneous graphs, (ii) construct an enhanced item–item subgraph from $R_2$, and (iii) propagate the signal into the UI graph through a lightweight matrix product.

\textbf{Construction of the Basic Heterogeneous Graph.} 
Given the user–item interaction data $\mathcal{D}_{UI}$ and the bundle–item affiliation data $\mathcal{D}_{BI}$, we construct
$G_{UI}=(\mathcal{U}\!\cup\!\mathcal{I},\,E_{UI})$ and $G_{BI}=(\mathcal{B}\!\cup\!\mathcal{I},\,E_{BI})$,
where $E_{UI}=\{(u,i)\mid (u,i)\in\mathcal{D}_{UI}\}$ and $E_{BI}=\{(b,i)\mid (b,i)\in\mathcal{D}_{BI}\}$.

\textbf{Cross-Domain Complementary Relation Injection.}
For each item pair $(i,j)$, compute the Jaccard similarities $J^{UI}_{i,j}$ and $J^{BI}_{i,j}$ on $G_{UI}$ and $G_{BI}$. We keep pairs that satisfy
$J^{UI}_{i,j}>\theta^{\mathrm{high}}_{UI}$ and $J^{BI}_{i,j}<\theta^{\mathrm{low}}_{BI}$ (i.e., $(i,j)\in R_2$), and define the enhanced item–item adjacency
\begin{equation}
A^{\mathrm{enh}}_{II}(i,j)=\mathbb{I}\{i=j\}+\mathbb{I}\{(i,j)\in R_2\}.
\label{eq:II_enh}
\end{equation}
Here $\mathbb{I}\{\cdot\}$ denotes the indicator function.

\textbf{Matrix-Multiplication–Driven Interaction Expansion.}
Let $A_{UI}$ be the UI adjacency of $G_{UI}$. The enhanced UI adjacency is obtained by
\begin{equation}
A'_{UI}=A_{UI}+\alpha\,\big(A_{UI}\cdot A^{\mathrm{enh}}_{II}\big),
\label{eq:UI_enh}
\end{equation}
where $\alpha\in[0,1]$ controls the injection strength. 

\subsection{Popularity-Aware Bundle Embedding Representation}
\quad In the cold-start bundle recommendation scenario, cold-start bias manifests itself mainly as inaccurate embedding representation of new bundles due to the lack of user interaction data. Thus, new bundles cannot be effectively represented, making it difficult to recommend. Therefore, we innovatively designed a bundle representation method based on item popularity, which essentially converts cold-start bundles into popular item representations, thereby breaking through the traditional model's strong dependence on UB interaction data.

\subsubsection{Item Popularity Quantification}
We propose a dual-source popularity measurement framework to capture both explicit user preference and implicit bundle-driven influence.

First, we define the \textbf{UI Popularity} as an item's global interaction frequency. For each item $i$, this is calculated as:
\begin{equation}
pop_i^{UI} = N_i^{UI} = \sum_{u \in \mathcal{U}} \mathbb{I}\{(u,i) \in \mathcal{D}_{UI}\},
\end{equation}
where $\mathcal{D}_{UI}$ represents the user-item interaction data.

Second, to mitigate cold-start bias, we compute the \textbf{Weighted BI Popularity}. This involves three steps: 
\textbf{(1) Bundle Popularity Normalization:} We calculate the user interaction count $N_b^{UB}$ for each bundle $b$:
\begin{equation}
N_b^{UB} = \sum_{u \in \mathcal{U}} \mathbb{I}\{(u,b) \in \mathcal{D}_{UB}\}.
\end{equation}
\textbf{(2) Percentile-based Normalization:} We then normalize this count to obtain a weight $w_b$ for each bundle:
\begin{equation}
w_b = \min\left( \frac{N_b^{UB}}{P_{90}}, 1 \right),
\end{equation}
where $P_{90}$ is the 90th percentile of bundle interaction counts.
\textbf{(3) Item BI Popularity Aggregation:} Finally, we aggregate these weights for all bundles affiliated with item $i$:
\begin{equation}
pop_i^{BI} = \sum_{b \in \mathcal{B}_i} w_b,
\end{equation}
where $\mathcal{B}_i = \{b | (b,i) \in \mathcal{D}_{BI}\}$ is the set of bundles containing item $i$.

Finally, the total popularity of an item is calculated as the sum of both components:
\begin{equation}
pop_i = pop_i^{UI} + pop_i^{BI}.
\end{equation}

\subsubsection{Long-Tail Adaptive Scaling}

To address the skewed popularity distribution of items, we design a piecewise scaling function. First, we compute the median value \(M\) of the item popularity \(pop_i\). This median-based segmentation divides the items into different percentile intervals, which are used to adjust the popularity scores.

The scaling strategy is based on the percentile distribution of \(pop_i\). Specifically, we compute the 50th, 60th, 70th, 80th, and 90th percentiles of \(pop_i\), denoted as \(P_{50}\), \(P_{60}\), \(P_{70}\), \(P_{80}\), and \(P_{90}\), respectively.

Next, we apply a piecewise linear scaling function to adjust the popularity as follows:
\begin{equation}
\begin{split}
pop_i^{\text{adj}} &= M \cdot \Bigg( 1 + \sum_{k=1}^5 \alpha_k \cdot \frac{pop_i - P_{L_k}}{P_{U_k} - P_{L_k}} \cdot \mathbb{I}\{P_{L_k} < pop_i \leq P_{U_k}\} \Bigg) \\
&  + M \cdot \mathbb{I}\{pop_i > P_{90}\}
\end{split}
\label{pop_i}
\end{equation}
where $\alpha_k$ is the interval adjustment coefficient. $P_{L_k}$ and $P_{U_k}$ are lower and upper bounds of $P$.

\subsubsection{Item Popularity-Aware Bundle-item Graph}

We inject item popularity information into the original bundle-item (BI) graph \(G_{BI}\) by updating the edge weights based on item popularity. For each bundle-item pair \((b, i)\), the edge weight is updated as follows:

\begin{equation}
w_{bi}^{pop} = \begin{cases} 
pop_i^{\text{adj}} & \text{if } \mathbb{I}((b,i)\in E_{BI}) \\
0 & \text{otherwise}
\end{cases}.
\end{equation}

The resulting item popularity-aware BI graph, denoted as \(\hat{G}_{BI}\), is a key component of our \textbf{cold-start scenario module} (detailed in Sec.~\ref{subsubsec:cold-scenario}). Its purpose is to generate bundle representations by aggregating the embeddings of their constituent items. Specifically, after obtaining the item embeddings from the enhanced User-Item view within the cold-start path (i.e., \(e_{i}^{UI,cold}\)), we compute the corresponding bundle embeddings for that view (\(e_b^{UI,cold}\)) using the following popularity-weighted aggregation:
\begin{equation}
\label{eq:final_pop_agg}
e_{b}^{UI,cold} = \sum_{i \in \mathcal{N}_b} \frac{w_{bi}^{pop} \cdot e_{i}^{UI,cold}}{\sqrt{D_b(b) \cdot D_i(i)}},
\end{equation}
where \(e_i^{UI,cold}\) represents the item embedding learned from the 'User-Item-Bundle View' of the cold-start module. 
The degree matrices \(D_b(b)\) and \(D_i(i)\) are defined as:
\begin{equation}
    D_b(b) = \sum_i w_{bi}^{pop}, \quad D_i(i) = \sum_b w_{bi}^{pop}.
\end{equation}
The full process for generating \(e_i^{UI,cold}\) is provided in Sec.~\ref{subsubsec:cold-scenario}. This calculation is crucial as it creates a rich representation for new bundles by leveraging item popularity and enhanced item relations, without relying on direct user-bundle interaction history. 

\subsection{Dual-Scenario Multi-View Contrastive Learning Framework}
\quad In previous work, models built for the warm-start scenario primarily focused on the interaction history between users and bundles, or constructed different views centered around users and bundles. This approach achieved good performance across multiple models and, by dynamically adjusting view weights, also demonstrated strong recommendation capabilities in the cold-start scenario. However, as pointed out by MultiCBR, the mixing of relations obtained from different views can interfere with each other. Thus the criteria for recommendations varies between the cold-start and  warm-start scenarios, if preference information from  different scenarios also interferes with each other. To address these limitations, we propose a dual-scenario framework that integrates both cold and warm scenarios.

In these two scenarios, we independently utilize different initial embeddings for users (\(e_{u}^{warm}, e_{u}^{cold}\)), items (\(e_{i}^{warm}, e_{i}^{cold}\)), and warm-start bundles (\(e_{b}^{warm}\)). 
Notably, there are no independent initial embeddings for cold-start bundles. 
This is because a core aspect of our approach is to derive the representation for a new bundle by aggregating the embeddings of its constituent items, which is fundamental to addressing the data scarcity issue for new bundles. Leveraging our core innovations, the model places particular emphasis on the cold-start scenario.

\subsubsection{Cold-start Scenario}
\label{subsubsec:cold-scenario}
The cold-start scenario module is specifically engineered to handle situations where user-bundle interaction data is sparse or non-existent for new bundles. To mitigate potential interference from the unreliable user-bundle graph in such cases, we exclusively utilize the enhanced interaction graphs derived from item-pair relations ($\hat{G}_{UI}$) and popularity-aware bundle-item affiliations ($\hat{G}_{BI}$). Through two symmetrically designed views, we construct a U-shaped information flow pathway from user to user, primarily guided by item-level relationships and popularity signals. 

%Furthermore, we incorporate a contrastive loss based on item-pair relations within the loss function to enhance learning in this scenario.

\textbf{1. User-Item-Bundle View:} This view aims to capture user preferences enriched by item-pair relations and subsequently generate bundle representations. Operating based on the enhanced UI graph $\hat{G}_{UI}$, bidirectional message propagation is performed using LightGCN. The message passing and aggregation process in LightGCN are defined as:
\begin{equation}
\label{LightGCN}
\left\{
\begin{aligned}
    e_{u}^{(t+1)} &= \sum_{i \in \mathcal{N}_u} \frac{e_{i}^{(t)}}{\sqrt{|\mathcal{N}_u| \cdot |\mathcal{N}_i|}}, \\
    e_{i}^{(t+1)} &= \sum_{u \in \mathcal{N}_i} \frac{e_{u}^{(t)}}{\sqrt{|\mathcal{N}_i| \cdot |\mathcal{N}_u|}}.
\end{aligned}
\right.
\end{equation}
Here, we apply this to the UI edges, using $e_{u}^{UI,cold(t)}$ and $e_{i}^{UI,cold(t)}$ to denote the $t$-th layer embeddings in this view. The final user and item embeddings \(e_{u}^{UI,cold}\) and \(e_{i}^{UI,cold}\) are derived through layer coalescence:
\begin{equation}
    e_{u}^{UI,cold} = \frac{1}{T} \sum_{t=0}^T e_{u}^{UI,cold(t)}, \quad e_{i}^{UI,cold} = \frac{1}{T} \sum_{t=0}^T e_{i}^{UI,cold(t)}.
\end{equation}
Subsequently, the bundle embedding representation $e_b^{UI,cold}$ is generated by  weighted aggregating item embeddings based on the enhanced BI graph, as shown in already detailed Equation~\ref{eq:final_pop_agg}. This step completes the first information-passing pathway of the cold-start module.

\textbf{2. Bundle-Item-User View:} Focusing on bundle composition information, this view processes the popularity-aware bundle-item graph $\hat{G}_{BI}$. Using $e_b^{UI,cold}$ and $e_{i}^{cold}$ as inputs, the LightGCN process (Equation~\ref{LightGCN}) is employed to obtain the aggregated representations $e_{b}^{BI,cold}$ and $e_{i}^{BI,cold}$. Information is then transmitted back to the user representations; user embeddings $e_u^{BI,cold}$ are generated through a aggregation leveraging the enhanced UI graph $\hat{G}_{UI}$,  using the following aggregation equation:

\begin{equation}
\label{aggregation}
    e_u^{BI,cold} = \frac{1}{|\mathcal{N}_i^{UI}|} \sum_{i \in \mathcal{N}_i^{UI}} e_{i}^{BI,cold}.
\end{equation}

\subsubsection{Warm-start Scenario}
For the warm-start scenario, where ample historical user-bundle interactions are available, the module processes information through three interaction views, building upon approaches similar to MultiCBR. This module utilizes the standard user-bundle interaction graph, the user-item interaction graph, and the popularity-enhanced interaction graph on the bundle-item affiliation view.

\textbf{1. User-Bundle Interaction View:} This view processes the user-bundle interaction graph using bidirectional LightGCN message propagation (Equation~\ref{LightGCN}) to obtain user and bundle embeddings ($e_{u}^{UB}$ and $e_{b}^{UB}$) through layer coalescence.

\textbf{2. User-Item Interaction View:} Based on the user-item interaction graph \(G_{UI}\), LightGCN propagation (Equation~\ref{LightGCN}) yields user and item embeddings \(e_{u}^{UI,warm}\) and \(e_{i}^{UI,warm}\). Bundle embeddings \(e_b^{UI,warm}\) are subsequently generated by aggregating item embeddings based on the popularity-aware BI graph \(\hat{G}_{BI}\).

\textbf{3. Bundle-Item Affiliation View:} Utilizing the popularity-aware BI graph \(\hat{G}_{BI}\), LightGCN propagation (Equation~\ref{LightGCN}) produces bundle and item embeddings \(e_b^{BI,warm}\) and \(e_{i}^{BI,warm}\). User embeddings \(e_u^{BI,warm}\) are derived through aggregation leveraging the UI graph \(G_{UI}\), following a process similar to Equation~\ref{aggregation}.

\begin{table*}[h!]
\centering
\caption{Summary of three real-world datasets where ``dens.'' denotes the density of a matrix.}
\label{tab:dataset_summary}
\begin{tabular}{l|rrrrrr|r}
\toprule
\textbf{Dataset} & \textbf{Users} & \textbf{Bundles} & \textbf{Items} & \textbf{User-bundle (dens.)} & \textbf{User-item (dens.)} & \textbf{Bundle-item (dens.)} & \textbf{Avg. size of bundle} \\
\midrule
Youshu$^1$ & 8,039 & 4,771 & 32,770 & 51,377 (0.13\%) & 138,515 (0.05\%) & 176,667 (0.11\%) & 37.03 \\
NetEase$^1$ & 18,528 & 22,864 & 123,628 & 302,303 (0.07\%) & 1,128,065 (0.05\%) & 1,778,838 (0.06\%) & 77.80 \\
iFashion$^1$ & 53,897 & 27,694 & 42,563 & 1,679,708 (0.11\%) & 2,290,645 (0.10\%) & 106,916 (0.01\%) & 3.86 \\
\bottomrule
\multicolumn{8}{l}{$^1$ \scriptsize{https://github.com/mysbupt/CrossCBR}} \\
\end{tabular}
\end{table*}

\subsubsection{Representation Fusion}
Utilizing the early fusion strategy\cite{ma2024multicbr}, embeddings of users and bundles generated from different views within each scenario are integrated, resulting in different scenarios embeddings $e_u^{warm}, e_u^{cold}, e_b^{warm}, e_b^{cold}$.

To flexibly integrate cold-start and warm-start scenarios, a scenario weight parameter, denoted as $k$, is introduced to control model bias. For user embeddings, the weighted concatenation of $e_u^{warm}$ from the warm-start scenario and $e_u^{cold}$ from the cold-start scenario is obtained using the following formula:
\begin{equation}
    e_u^{all}=(e_u^{warm} \cdot k) \oplus (e_u^{cold} \cdot (1 - k)),
\end{equation}
where $\oplus$ denotes the concatenate operation. Similarly, the embedding of the bundled package undergoes the same operation:
\begin{equation}
    e_b^{all}=(e_b^{warm} \cdot k) \oplus (e_b^{cold} \cdot (1 - k)).
\end{equation}
This achieves an effective integration of the two scene embeddings.

\subsubsection{Joint Optimization}
In this work, we innovatively propose a dual-scenario contrastive loss and an item-pair relations enhanced contrastive loss. While retaining the user-bundle contrastive loss, we directly adjust the item embeddings based on item-pair relations, and utilize the contrastive loss between scenarios to achieve information synergy.

In the user-bundle contrastive loss, we continue to utilize the self-supervised contrastive framework\cite{ma2024multicbr}. We generate positive sample pairs through data augmentation methods involving noise injection, construct negative pairs using within-batch negative sampling, and employ InfoNCE. The equations are as follows:
%用下面的吧 上面的看起来太密了

\begin{equation}
\left\{
\begin{aligned}
    \mathcal{L}_{u}^C &= \frac{1}{|\mathcal{U}|} \sum_{u \in \mathcal{U}} -\log \frac{\exp(\cos(e_u', e_u'')/\tau)}{\sum_{v \in \mathcal{U}} \exp(\cos(e_u', e_v'')/\tau)}, \\
    \mathcal{L}_{b}^C &= \frac{1}{|\mathcal{B}|} \sum_{b \in \mathcal{B}} -\log \frac{\exp(\cos(e_b', e_b'')/\tau)}{\sum_{m \in \mathcal{B}} \exp(\cos(e_b', e_m'')/\tau)}.
\end{aligned}
\right.
\end{equation}
where $e_u', e_u'', e_b', e_b''$ represent the positive sample pairs of user $u$ and bundle $b$ generated through two distinct noise injections, $\mathcal{U}$ and $\mathcal{B}$ denote the current batch user set and bundle set, respectively, and $\tau$ is the temperature coefficient that controls the smoothness of the similarity distribution.

In the dual-scenario contrastive loss, we construct the contrastive loss by embedding the user and bundle in two scenarios respectively as $e^w_u, e^c_u, e^w_b, e^c_b$. The positive samples are composed of the same user and bundle, using InfoNCE as the contrastive loss function. The equation is as follows:
\begin{equation}
\begin{aligned}
\mathcal{L}^{scen} &= \frac{1}{|\mathcal{U}|} \sum_{u \in \mathcal{U}} -\log \frac{\exp(\cos(e^w_u, e^c_u)/\tau)}{\sum_{v \in \mathcal{U}} \exp(\cos(e^w_u, e^c_v)/\tau)} \\
&+ \frac{1}{|\mathcal{B}|} \sum_{b \in \mathcal{B}} -\log \frac{\exp(\cos(e^w_b, e^c_b)/\tau)}{\sum_{m \in \mathcal{B}} \exp(\cos(e^w_b, e^c_m)/\tau)}
\end{aligned}
\end{equation}

In the context of item-pair relations modeling, we propose a item-pair contrastive loss, which enhances the embedding representation by explicitly distinguishing between complementary items and irrelevant items. Specifically, for the embedding $e_i$ of item $i$, its positive samples consist solely of its own embedding $e_i$, while negative samples are selected from the set of negative samples $\mathcal{N}^{R_4}_i$ (where $R_4$ denotes the negative relation). The contrastive loss function employs InfoNCE, and the mathematical expression is as follows:
\begin{equation}
    \mathcal{L}_{i}^C = -\frac{1}{|\mathcal{I}|} \sum_{i \in \mathcal{I}} \log \frac{ \exp(1/\tau) }{ \exp(1/\tau) + \sum_{k \in \mathcal{N}_i^{R4}} \exp( \cos(e_i, e_k)/\tau )}.\label{eq:iiloss}
\end{equation}

BPR loss has been widely used as a part of loss function in contemporary mainstream bundle recommendation models, with its mathematical formula as follows:
\begin{equation}
    \mathcal{L}^{BPR} = \sum_{(u, b, b') \in Q} -\ln\sigma(y^*_{u, b} - y^*_{u, b'}),
\end{equation}
where \(Q = \{(u, b, b')|u \in \mathcal{U}, b, b' \in \mathcal{B}, x_{ub} = 1, x_{ub'} = 0\}\), and \(\sigma(\cdot)\) is the sigmoid function.

To generate the preference score \(y^*_{u,b}\) for user \(u\) on bundle \(b\), we compute the inner-product between the overall user and bundle representations, as follows:
\begin{equation}
    y^*_{u,b} = e_u \cdot e_b,
\end{equation}
where \(e_u\) is the embedding of user \(u\) and \(e_b\) is the embedding of bundle \(b\). The score \(y^*_{u,b'}\) is similarly computed for the negative sample bundle \(b'\). 
%The BPR loss function then optimizes the model to maximize the difference between the positive and negative preference scores for a given user-bundle pair.

Finally, by integrating the BPR loss with three comparative loss terms, the loss function $\mathcal{L}$ of our model is constructed as follows:
\begin{equation}
    \mathcal{L} = \mathcal{L}^{BPR} + \beta_1 (\mathcal{L}_{u}^C+\mathcal{L}_{b}^C) +\beta_2 \mathcal{L}^{scen} + \beta_3 \mathcal{L}_{ii}^C + \beta_4 \|\Theta\|_2^2,
\end{equation}
where $\beta_1, \beta_2, \beta_3, \beta_4$ are hyperparameters used to balance different loss terms. $\|\Theta\|_2^2$ represents the regularization term.
% $\Theta = E^w_U, E^w_I, E^w_B$\newline$, E^c_U, E^c_I$.

%In particular, in the cold start scenario, we increased $\beta_1$ to pay more attention to the similarity between items rather than the probability of being selected.

\section{Experiments}

In the experiment, we will address the following questions:
%\vspace*{-\baselineskip}  % 移除一个行高的空白
\begin{enumerate}[label={RQ\arabic*.}] % 自定义标签
    \item How does our model perform in cold-start bundle recommendation compared to existing state-of-the-art methods?
    \item How does our model perform in warm-start bundle recommendation compared to existing state-of-the-art methods.
    \item What is the contribution of each component of our model to its effectiveness in the ablation study?
    \item Is the dual-scenario contrastive learning framework effective in improving the performance of our model?
\end{enumerate}

\setlength{\tabcolsep}{3pt}
\begin{table*}[ht]
\setlength{\abovecaptionskip}{0pt}
\centering
\caption{Performance comparison of EpicCBR and baseline cold-start methods on three real-world datasets.}
\resizebox{\textwidth}{!}{
\begin{tabular}{l|ccc|ccc|ccc|ccc|ccc|ccc}
\toprule
\multirow{2}{*}{Model} & \multicolumn{6}{c|}{Youshu} & \multicolumn{6}{c|}{NetEase} & \multicolumn{6}{c}{iFashion} \\
\cmidrule(lr){2-4} \cmidrule(lr){5-7} \cmidrule(lr){8-10} \cmidrule(lr){11-13} \cmidrule(lr){14-16} \cmidrule(lr){17-19}
 & \multicolumn{3}{c|}{Recall@20} & \multicolumn{3}{c|}{nDCG@20} & \multicolumn{3}{c|}{Recall@20} & \multicolumn{3}{c|}{nDCG@20} & \multicolumn{3}{c|}{Recall@20} & \multicolumn{3}{c}{nDCG@20} \\
 & Cold & Warm &All & Cold & Warm & All & Cold & Warm & All & Cold & Warm & All & Cold & Warm &All & Cold & Warm & All \\
\midrule
DropoutNet\cite{volkovs2017dropoutnet} &.0022 &.0336 &.0148 &.0007 &.0153 &.0055 &.0028 &.0154 &.0046 &.0015 &.0078 &.0024 &.0009 &.0060 &.0039 &.0008 &.0045 &.0027 \\
CVAR\cite{zhao2022improving} &.0008 &.1958 &.0829 &.0002 &.1112 &.0533 &.0002 &.0308 &.0156 &.0001 &.0154 &.0084 &.0007 &.0220 &.0125 &.0004 &.0015 &.0084 \\
CLCRec\cite{wei2021contrastive} &.0137 &.0626 &.0367 &.0087 &.0317 &.0194 &.0136 &.0407 &.0259 &.0075 &.0215 &.0138 &.0053 &.0203 &.0126 &.0043 &.0013 &.0085 \\
CCFCRec\cite{zhou2023contrastive} &.0044 &.1554 &.0702 &.0022 &.0798 &.0425 &.0007 &.0265 &.0130 &.0004 &.0128 &.0068 &.0005 &.0439 &.0252 &.0003 &.0030 &.0172 \\
CoHeat\cite{jeon2024cold} &\underline{.0183} &.2804 &.1247 &\underline{.0105} &.1646 &.0833 &\underline{.0191} &.0847 &.0453 &\underline{.0093} &.0455 &.0264 &\underline{.0170} &.1156 &.0658 &\underline{.0096} &z.0087 &.0504 \\
MultiCBR\cite{ma2024multicbr} &.0003 &\underline{.2855} &\underline{.1287} &.0001 &\underline{.1680} &\underline{.0857} &.0000 &\underline{.0919} &\underline{.0488} &.0000 &\underline{.0498} &\underline{.0284} &.0006 &\textbf{.1504} &\underline{.0871}  &.0003  &\textbf{.1207} &\underline{.0703}\\
% PET \cite{kim2024towards}  &.0 &.0 &.0 &.0 &.0 &.0 &.0 &.0 &.0 &.0 &.0 &.0 &.0 &.0 &.0 &.0 &.0 &.0 \\
\midrule
EpicCBR (ours) &\textbf{.0393} &\textbf{.2896} &\textbf{.1329} &\textbf{.0288} &\textbf{.1722} &\textbf{.0884} &\textbf{.0399} &\textbf{.0931} &\textbf{.0497} &\textbf{.0224} &\textbf{.0503} &\textbf{.0288} &\textbf{.0828} &\underline{.1474} &\textbf{.1206} &\textbf{.0711} &\underline{.1189} &\textbf{.1043} \\
\bottomrule
\end{tabular}
  }
\label{tab:cold_start_performance}
\end{table*}

% \vspace*{-\baselineskip} 

\subsection{Experiment Setup}
\quad \textbf{Datasets.} To evaluate our model, we conducted experiments on three real-world datasets, which are popular in bundle recommendation, as summarized in Table \ref{tab:dataset_summary}. The datasets related to books consists of book lists from a book review website. The Netease Cloud dataset contains playlists from an online music platform. And the ifashion dataset consists of bundled items from a shopping website.

\textbf{Design and Utilization of Co-occurrence Matrix.}
In view enhancement, experiments using both similar and complementary pairs was conducted. Although including similar pairs led to a substantial increase in training time overhead, the model performance did not improve significantly. Thus, we opted for complementary pairs for view augmentation, hypothesizing that similar pairs may carry less informative value.

During the contrastive learning enhancement process, various methods were employed to compare positive and negative pairs. The experiments revealed that retaining only the positive pairs did not enhance performance, whereas using negative pairs derived from the co-occurrence matrix significantly improved the contrastive loss outcomes.

\textbf{Baseline Cold-Start Methods.}  
We compared our model with Coheat\cite{jeon2024cold}, the state-of-the-art (SOTA) model for cold-start bundle recommendation. Coheat employs a popularity-based fusion method to address the skewed distribution of user-bundle interactions. Additionally, it utilizes curriculum learning and contrastive learning to effectively learn latent representations. We also compared our model with existing cold-start item recommendation methods, which can learn from bundle-item information by treating bundles as items for recommendation.  CVAR\cite{zhao2022improving} is a model-agnostic recommendation method based on autoencoders, which pre-warms the IDs of new items. CLCRec\cite{wei2021contrastive} and CCFCRec\cite{zhou2023contrastive} use contrastive learning techniques to address the cold-start problem. For these methods, we utilized bundle-item multi-hot vectors as content information, following the contrastive approach presented in the Coheat paper.  Furthermore, we compared our model with the SOTA warm-start model, MultiCBR\cite{ma2024multicbr}, to demonstrate its adaptability in both cold-start and warm-start scenarios. The experimental results of the aforementioned models were calibrated using Coheat's experimental framework.

\textbf{Baseline warm-start methods.} To evaluate our model's performance in warm scenarios, we compared our model with the existing warm-start models.  MultiCBR\cite{ma2024multicbr} leverages multiple collaborative filtering models to capture diverse user-item interactions, combining their strengths to improve recommendation accuracy and robustness. BGCN\cite{chang2020bundle} and CrossCBR\cite{ma2022crosscbr} are other models that perform bundle recommendation using graph neural network. LightGCN\cite{he2020lightgcn}, MFBPR\cite{rendle2012bpr} use matrix factorization and graph learning to perform item recommendation respectively. We also compared our model with existing contrastive models, such as SGL \cite{wu2021self}, SimGCL\cite{yu2022graph}, and LightGCL\cite{cai2023lightgcl}. Coheat\cite{jeon2024cold} uses popularity based strategy, which is also relevant to our study.

\textbf{Evaluation metrics.} We adopt Recall@k and nDCG@k as metrics to evaluate the performance of our model. Recall@k examines the accuracy of the recommendations among the top k recommended items. In contrast, nDCG@k evaluates the model's performance by quantifying the disparity between the obtained ranking and the ideal ranking.

\textbf{Experimental process.} We conduct experiments in the cold-start scenario. The dataset is divided into subsets in a ratio of 7:1:2 for training, validation, and testing purposes. In the cold-start scenario, the bundles in the test set have not appeared in the training set, ensuring that they are completely cold-start for the model. We report the best Recall@20 and nDCG@20 obtained as the basis evaluation.

\textbf{Hyperparameter Selection and Model Configuration.}  
We utilize the baselines with their official implementations and use the reported best hyperparameters. For methods that did not test on the same datasets as ours, we perform grid search for key parameters to determine the optimal settings. For our model, we conduct grid search on each dataset for the following hyperparameters: the enhancement edge weight \(\alpha \in \{0, 0.1, 0.2, \dots, 1.0\}\), the scenario weight \(k \in \{0.1, 0.2, \dots, 1.0\}\), the temperature parameters \(\tau \in \{0.1, 0.2, \dots, 1.0\}\), and the loss coefficients \(\beta_1, \beta_2, \beta_3, \beta_4 \in [0, 1]\). This approach ensures that the model parameters are carefully tuned for each dataset. In the cold-start scenario, we increase $\beta_1$ parameter to place greater emphasis on the similarity signals mined from user and bundle pairs, thereby reducing reliance on historical user-bundle interactions.

\subsection{Comparison with Cold-start Methods (RQ1)}
\quad In Table~\ref{tab:cold_start_performance}, We conducted comprehensive comparisons between our model and state-of-the-art cold-start methods across multiple datasets. Experimental results demonstrate that our model achieves significant performance improvements over existing approaches in all evaluated scenarios, validating its superior effectiveness. Notably, on the cold-start iFashion dataset, our model surpasses the current SOTA model for this task by 387\% in Recall@20, highlighting its remarkable advantage in handling extreme cold-start scenarios.

\textbf{Item‐Pair Relation Modeling + Popularity‐Aware Bundle Representation.}
We jointly exploit user–item interaction signals and bundle preference patterns to learn explicit item–item relations, while the Popularity component assigns larger weights to items that are selected more frequently within bundles.  
Unlike MultiCBR, which builds UI and BI graphs solely from historical interactions, EpicCBR predicts latent relations between items unseen by the user, yielding more precise user portraits; the popularity weighting further injects strong signals from hot items, providing critical clues for recommending cold‐start bundles to potential users.

\textbf{Why does \textit{iFashion} gain the most?}
With an average bundle size of only 3–4 items, iFashion exhibits extreme BI sparsity.  
The model therefore becomes highly sensitive to the quality of each single item’s representation.  
Our relation modeling allows high‐interaction yet low‐co‐occurrence item pairs (popular but rarely co‐bundled) to complement each other’s information, while popularity weighting amplifies the features of these key items, effectively mitigating information loss caused by sparsity and leading to the largest performance improvement on iFashion.

\textbf{Item‐Pair Contrastive Loss.}
Building on the learned relations, we impose a item-relation contrastive loss that directly pulls semantically related item pairs together and pushes unrelated pairs apart, refining embeddings under cold‐start conditions.  
The loss simultaneously preserves global structural similarity and local interaction semantics, aligning the positions of popular representative items and their complementary or similar partners in the embedding space with the true underlying relations.   
Together, relation modeling with popularity weighting provides discriminative structural priors, while the contrastive loss sharpens semantic boundaries, jointly underpinning the strong cold‐start performance of EpicCBR.

\begin{table}[t]
    \centering
    \setlength{\abovecaptionskip}{0pt}  % 去掉标题上方的间距
    \caption{Experiments for warm-start dataset}
    \label{tab:warm start}
    \setlength{\tabcolsep}{2pt} % Adjust column separation for tighter spacing
    \begin{tabular}{l|cc|cc|cc}
        \toprule
        Model & \multicolumn{2}{c}{Youshu} & \multicolumn{2}{c}{NetEase} & \multicolumn{2}{c}{iFashion} \\
        \cmidrule(lr){2-3} \cmidrule(lr){4-5} \cmidrule(lr){6-7}
        & \makecell{Recall\\@20} & \makecell{nDCG\\@20} & \makecell{Recall\\@20} & \makecell{nDCG\\@20} & \makecell{Recall\\@20} & \makecell{nDCG\\@20} \\
        \midrule
            MFBPR\cite{rendle2012bpr}& .1959 & .1117 & .0355 & .0181 & .0752 & .0542 \\
            LightGCN\cite{he2020lightgcn}& .2286 & .1344 & .0496 & .0254 & .0837 & .0612 \\
            SGL\cite{wu2021self}& .2568 & .1527 & .0687 & .0368 & .0933 & .0690 \\
            SimGCL\cite{yu2022graph}& .2691 & .1593 & .0710 & .0377 & .0919 & .0677 \\
            LightGCL\cite{cai2023lightgcl}& .2712 & .1607 & .0722 & .0388 & .0943 & .0686 \\
        \midrule
            BGCN\cite{chang2020bundle}& .2347 & .1345 & .0491 & .0258 & .0733 & .0531 \\
            CrossCBR\cite{ma2022crosscbr}& .2776 & .1641 & .0791 & .0433 & .1133 & .0875 \\
            CoHEAT\cite{jeon2024cold}& .2804 & .1646 & .0847 & .0455 & .1156 & .0876 \\
        \midrule
        EpicCBR (ours) & \textbf{.2896} & \textbf{.1722} & \textbf{.0931} & \textbf{.0503} & \textbf{.1474} & \textbf{.1189} \\
        \bottomrule
    \end{tabular}
\vspace*{-\baselineskip}  % 移除一个行高的空白
\end{table}

\subsection{Comparison with Warm-start Methods(RQ2)}
\quad To assess EpicCBR's adaptability in data-rich scenarios, we compare it with representative warm-start bundle recommenders in Table~\ref{tab:warm start}. Graph-based contrastive methods like SGL~\cite{wu2021self}, SimGCL~\cite{yu2022graph}, and LightGCL~\cite{cai2023lightgcl}, as well as methods leveraging structural biases and popularity cues (e.g., BGCN~\cite{chang2020bundle}, CoHeat~\cite{jeon2024cold}), demonstrate strong performance, highlighting the value of discriminative item embeddings and structural information under dense user-bundle links.

Operating similarly to a popularity-enhanced MultiCBR in warm-start, EpicCBR is expected to converge in performance with abundant data due to structural similarities. Table~\ref{tab:warm start} largely aligns with this, yet reveals dataset-specific nuances. On Youshu and NetEase, the added popularity signal remains beneficial (Recall@20 gains of $\sim$2.1\% and $\sim$1.7\%), as larger bundles provide a richer set of BI edges where popularity can effectively distinguish connection importance. Conversely, a slight dip on iFashion ($\bar{\text{bundle size}} = 3.86$) suggests that with smaller bundles, supplementary popularity information can introduce noise, increasing sensitivity to individual items.

\begin{figure}[t] % 图片浮动体
    \centering
    \includegraphics[width=\columnwidth,keepaspectratio]{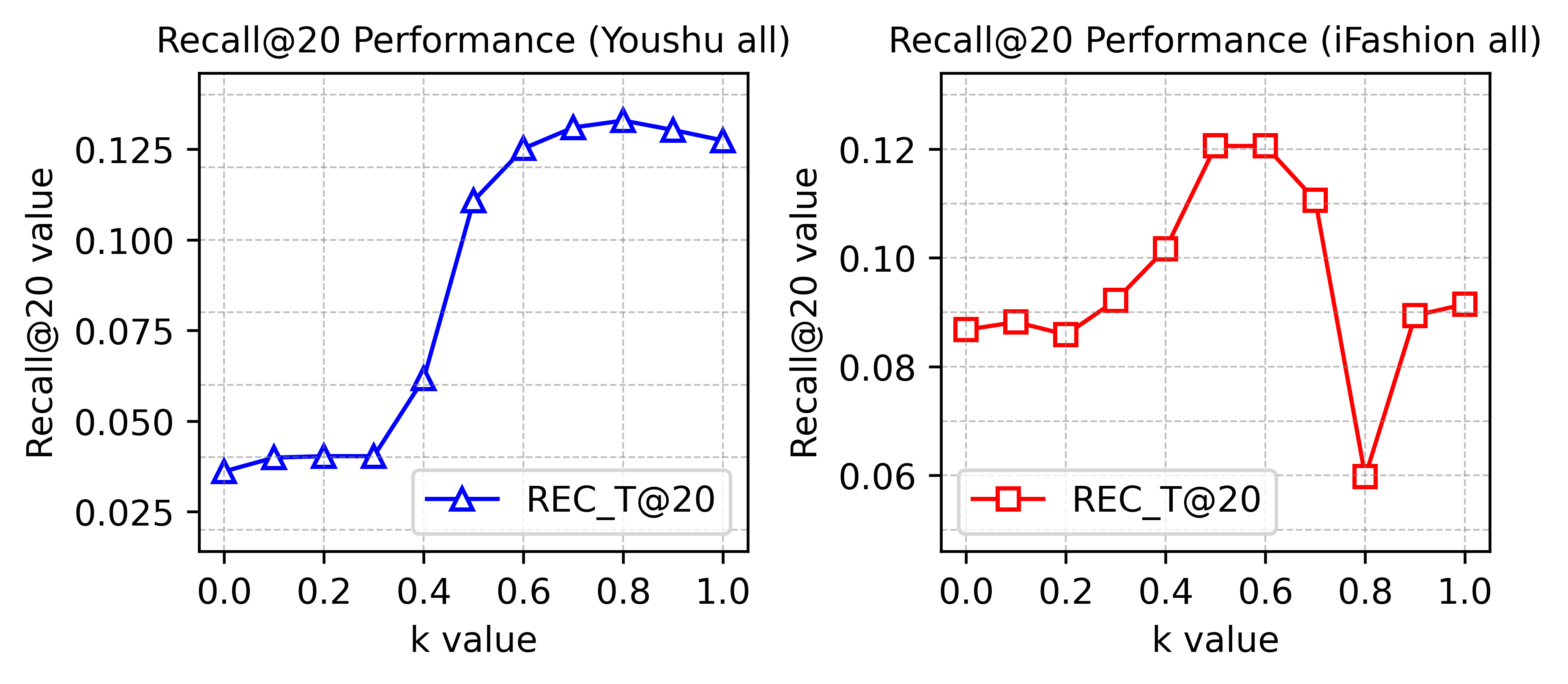} % 设置图片宽度为一栏宽度，并保持纵横比
    \caption{Experiments for scenario weight on dataset all} % 图片标题
    \label{fig:scenario_weight} % 图片标签
\end{figure}

\subsection{Ablation Study (RQ3)}
\quad The  Table~\ref{tab:ablation} presents an ablation study comparing our proposed model against four variants:EpicCBR-enh, EpicCBR-UI-enh, EpicCBR-pop, and EpicCBR-iiloss. These investigations were carried out within a cold-start scenario. EpicCBR-enh use the same graph as MultiCBR to compare our model's performance with the original graph. In EpicCBR-UI-enh, the view enhancement in Equation~\ref{eq:UI_enh} was replaced by the original UI graph. For EpicCBR-pop, we expose the usage of popularity based view of Equation~\ref{eq:final_pop_agg}. The original BI graph was utilized for graph aggregation and propagation computations. In EpicCBR-iiloss the contrastive loss function of the item pairs shown in Equation~\ref{eq:iiloss} was deactivated. As demonstrated in the table, our model consistently outperforms all variants, indicating that all incorporated improvements contribute to the enhanced performance of the model. Notably, iiloss showed poor performance in the iFashion dataset, which might indicate that the item-based loss function could contribute negatively in small bundle datasets.

\begin{table}[t]
    \centering
    \setlength{\abovecaptionskip}{0pt}  % 去掉标题上方的间距
    \caption{Experiments for Ablation study.}
    \label{tab:ablation}
    \setlength{\tabcolsep}{2pt} % Adjust column separation for tighter spacing
    \begin{tabular}{l|cc|cc|cc}
        \toprule
        Model & \multicolumn{2}{c}{Youshu} & \multicolumn{2}{c}{NetEase} & \multicolumn{2}{c}{iFashion} \\
        \cmidrule(lr){2-3} \cmidrule(lr){4-5} \cmidrule(lr){6-7}
        & \makecell{Recall\\@20} & \makecell{nDCG\\@20} & \makecell{Recall\\@20} & \makecell{nDCG\\@20} & \makecell{Recall\\@20} & \makecell{nDCG\\@20} \\
        \midrule
        EpicCBR-enh & .0370 & .0194 & .0340 & .0186 & .0441 & .0295\\
        EpicCBR-UI-enh & .0381 & .0206 & .0372 & .0205 & .0741 & .0575\\
        EpicCBR-pop & .0321 & .0180 & .0355 & .0203 & .0471 & .0362\\
        EpicCBR-iiloss & \underline{.0385} & \underline{.0216} & \underline{.0381} & \underline{.0216} & \underline{.0764} & \underline{.0695} \\
        \midrule
        EpicCBR (ours) & \textbf{.0393} & \textbf{.0288} & \textbf{.0399} & \textbf{.0224} & \textbf{.0828} & \textbf{.0711} \\
        \bottomrule
    \end{tabular}
\vspace*{-\baselineskip}  % 移除一个行高的空白
\end{table}

\subsection{Dual-Scenario Contrastive Learning Framework Validation (RQ4)}

\quad To evaluate the efficacy of the proposed dual-scenario contrastive learning framework, experiments were conducted on the \emph{all} dataset split, which comprises both cold-start and warm-start recommendation instances. Evaluating performance on this mixed dataset is crucial, as it faithfully emulates a real-world online recommendation environment where diverse scenarios coexist. This allows us to effectively observe how the framework coordinates and leverages information from both cold and warm scenarios. Figure~\ref{fig:scenario_weight} shows the performance curves on the iFashion and Youshu datasets as a function of the scenario weight $k$. The results clearly indicate that this weight is a critical hyperparameter with a substantial impact on performance; fine-tuning $k$ is essential for maximizing accuracy. Varying $k$ allows us to control the balance between the cold-start and warm-start components of the model. Increasing the cold-start weight strengthens the reliance on information sources tailored for new bundles, while increasing the warm-start weight enhances the exploitation of abundant historical interaction data. The observed peak performance demonstrates that striking the right balance effectively combines the strengths of both scenarios, leading to improved overall results. This empirical evidence confirms that, under mixed conditions, the dual-scenario contrastive learning framework, when properly balanced, significantly contributes to robust and improved recommendation performance.

\section{Related work}
\subsection{Bundle Recommendation}

\quad Bundle recommendation\cite{sun2024survey} has been a subject of extensive research, primarily categorized into discriminative and generative tasks. Discriminative methods focus on predicting user preferences for existing bundles via modeling user-bundle interactions, while generative methods aim to generate novel bundles based on historical interactions. Our work primarily addresses the discriminative task, but its ability to uncover latent item relations (R1-R4, as defined in Sec.~\ref{subsubsec:relation_def}) also shows potential for generative tasks, as these relations can guide the design of appealing item combinations.

Existing discriminative methods leverage graph neural networks (GNNs) and contrastive learning to model interactions. For example, BGCN\cite{chang2020bundle} and CrossCBR\cite{ma2022crosscbr} use graph structures to capture user-bundle and bundle-item relations, but they rely heavily on historical user-bundle interactions, struggling with cold-start scenarios where such data is absent. MultiCBR\cite{ma2024multicbr} introduces multi-view contrastive learning for warm-start settings, but its mixed relations from different views may interfere with each other, especially when new bundles lack interaction signals. In contrast, our EpicCBR explicitly separates cold and warm scenarios, using item-relation-enhanced views (e.g., injecting R2 edges into UI graphs) to mitigate interference and improve robustness across scenarios.

Generative bundle recommendation methods, which focus on creating new bundles, often rely on item co-occurrence patterns. Our work complements these efforts by mining fine-grained item relations (e.g., complementary R1 and cross-domain R2), which may inform the generation of semantically consistent or user-preferred bundles. 

\subsection{Cold-Start Problem}
\quad The cold-start problem is crucial in recommendation systems as it directly impacts the ability to provide accurate suggestions for new users or items with limited interaction data\cite{volkovs2017dropoutnet,lee2019melu}. And cold-start problem\cite{wei2021contrastive} in bundle recommendation pertains to recommending bundles that have no prior interaction history with users. Existing approaches, such as Coheat\cite{jeon2024cold}, still rely on historical user-bundle interactions. In contrast, our work explores the underlying recommendation logic for cold-start bundles by leveraging user-item and bundle-item relations, thereby reducing the reliance on historical user-bundle interactions. Experimental results demonstrate that our method consistently outperforms current state-of-the-art approaches in this challenging scenario. Additionally, some studies leverage LLMs to tackle cold-start recommendations. Due to their inherent knowledge, LLMs excel in cold-start and zero-shot scenarios\cite{rong2024llm,sun2024revisiting}.
\subsection{Co-occurence Matrix}
\quad With reference to Nguyen’s work \cite{nguyen2024bundle}, we compute item co-occurrence matrices by multiplying the UI matrix and the BI matrix with their respective transposes. To the best of our knowledge, our study is the first to employ the Jaccard similarity coefficient on these co-occurrence matrices and to jointly analyze the relations between items derived from both the I-U-I and I-B-I structures. By categorizing item relations into four distinct types, we achieve a more granular and nuanced understanding of item pair associations. The four types we define can not only be applied to enhance information propagation paths in graph structures, but can also be used to design contrastive losses for adjusting item embeddings and guide future work in generating new bundles.

\section{Conclusion}

\quad In this paper, we propose EpicCBR, a novel approach addressing this issue through a dual-scenario framework designed for robustness across cold and warm settings. EpicCBR's core innovation lies in its deep utilization of item-level information: it mines four types of item relations to enhance user profiling, perform data augmentation, and guide item embedding learning, while also employing a popularity-based method for bundle representation, crucial for new bundles. Built upon a multi-view graph contrastive learning framework, EpicCBR effectively integrates diverse information. Extensive experiments on multiple real-world datasets demonstrate that EpicCBR significantly outperforms existing state-of-the-art (SOTA) methods in recommendation accuracy. Our method offers a new perspective on cold-start bundle recommendation and shows potential for designing popular bundles by leveraging intrinsic item connections. For future work, we plan to further explore the impact of mining intrinsic item relations on recommendation systems, and investigate the potential of applying the proposed dual-scenario framework to other recommendation domains.

\begin{acks}
This work was supported in part by the National Natural Science Foundation of China under Grant No. 62276110, No. 62172039 and in part by the fund of Joint Laboratory of HUST and Pingan Property \& Casualty Research (HPL). The authors would also like to thank the anonymous reviewers for their comments on improving the quality of this paper.
\end{acks}

\bibliographystyle{ACM-Reference-Format}
\balance
\bibliography{reference}

@inproceedings{chang2020bundle,
  title={Bundle recommendation with graph convolutional networks},
  author={Chang, Jianxin and Gao, Chen and He, Xiangnan and Jin, Depeng and Li, Yong},
  booktitle={Proceedings of the 43rd international ACM SIGIR conference on Research and development in Information Retrieval},
  pages={1673--1676},
  year={2020}
}

@article{sun2024survey,
  title={A Survey on Bundle Recommendation: Methods, Applications, and Challenges},
  author={Sun, Meng and Li, Lin and Li, Ming and Tao, Xiaohui and Zhang, Dong and Wang, Peipei and Huang, Jimmy Xiangji},
  journal={arXiv preprint arXiv:2411.00341},
  year={2024}
}

@inproceedings{kim2024towards,
  title={Towards Better Utilization of Multiple Views for Bundle Recommendation},
  author={Kim, Kyungho and Kim, Sunwoo and Lee, Geon and Shin, Kijung},
  booktitle={Proceedings of the 33rd ACM International Conference on Information and Knowledge Management},
  pages={3827--3831},
  year={2024}
}

@article{ma2024multicbr,
  title={Multicbr: Multi-view contrastive learning for bundle recommendation},
  author={Ma, Yunshan and He, Yingzhi and Wang, Xiang and Wei, Yinwei and Du, Xiaoyu and Fu, Yuyangzi and Chua, Tat-Seng},
  journal={ACM Transactions on Information Systems},
  volume={42},
  number={4},
  pages={1--23},
  year={2024},
  publisher={ACM New York, NY}
}

@inproceedings{jeon2024cold,
  title={Cold-start Bundle Recommendation via Popularity-based Coalescence and Curriculum Heating},
  author={Jeon, Hyunsik and Lee, Jong-eun and Yun, Jeongin and Kang, U},
  booktitle={Proceedings of the ACM Web Conference 2024},
  pages={3277--3286},
  year={2024}
}

@inproceedings{zhao2022improving,
  title={Improving item cold-start recommendation via model-agnostic conditional variational autoencoder},
  author={Zhao, Xu and Ren, Yi and Du, Ying and Zhang, Shenzheng and Wang, Nian},
  booktitle={Proceedings of the 45th International ACM SIGIR Conference on Research and Development in Information Retrieval},
  pages={2595--2600},
  year={2022}
}

@inproceedings{zhou2023contrastive,
  title={Contrastive collaborative filtering for cold-start item recommendation},
  author={Zhou, Zhihui and Zhang, Lilin and Yang, Ning},
  booktitle={Proceedings of the ACM Web Conference 2023},
  pages={928--937},
  year={2023}
}

@inproceedings{wei2021contrastive,
  title={Contrastive learning for cold-start recommendation},
  author={Wei, Yinwei and Wang, Xiang and Li, Qi and Nie, Liqiang and Li, Yan and Li, Xuanping and Chua, Tat-Seng},
  booktitle={Proceedings of the 29th ACM international conference on multimedia},
  pages={5382--5390},
  year={2021}
}

@inproceedings{sun2022revisiting,
  title={Revisiting bundle recommendation: datasets, tasks, challenges and opportunities for intent-aware product bundling},
  author={Sun, Zhu and Yang, Jie and Feng, Kaidong and Fang, Hui and Qu, Xinghua and Ong, Yew Soon},
  booktitle={Proceedings of the 45th International ACM SIGIR Conference on Research and Development in Information Retrieval},
  pages={2900--2911},
  year={2022}
}

@article{du2023enhancing,
  title={Enhancing item-level bundle representation for bundle recommendation},
  author={Du, Xiaoyu and Qian, Kun and Ma, Yunshan and Xiang, Xinguang},
  journal={ACM Transactions on Recommender Systems},
  year={2023},
  publisher={ACM New York, NY}
}

@article{liu2017modeling,
  title={Modeling buying motives for personalized product bundle recommendation},
  author={Liu, Guannan and Fu, Yanjie and Chen, Guoqing and Xiong, Hui and Chen, Can},
  journal={ACM Transactions on Knowledge Discovery from Data (TKDD)},
  volume={11},
  number={3},
  pages={1--26},
  year={2017},
  publisher={ACM New York, NY, USA}
}

@inproceedings{nguyen2024bundle,
  title={Bundle Recommendation with Item-Level Causation-Enhanced Multi-view Learning},
  author={Nguyen, Huy-Son and Bui, Tuan-Nghia and Nguyen, Long-Hai and Hoang, Hung and Thi Nguyen, Cam-Van and Le, Hoang-Quynh and Le, Duc-Trong},
  booktitle={Joint European Conference on Machine Learning and Knowledge Discovery in Databases},
  pages={324--341},
  year={2024},
  organization={Springer}
}

@article{jang2024attention,
  title={Attention-based multi attribute matrix factorization for enhanced recommendation performance},
  author={Jang, Dongsoo and Li, Qinglong and Lee, Chaeyoung and Kim, Jaekyeong},
  journal={Information Systems},
  volume={121},
  pages={102334},
  year={2024},
  publisher={Elsevier}
}

@inproceedings{wang2024relative,
  title={Relative Contrastive Learning for Sequential Recommendation with Similarity-based Positive Sample Selection},
  author={Wang, Zhikai and Shen, Yanyan and Zhang, Zexi and He, Li and Li, Yichun and Gu, Hao and Zhang, Yinghua},
  booktitle={Proceedings of the 33rd ACM International Conference on Information and Knowledge Management},
  pages={2493--2502},
  year={2024}
}

@inproceedings{he2020lightgcn,
  title={Lightgcn: Simplifying and powering graph convolution network for recommendation},
  author={He, Xiangnan and Deng, Kuan and Wang, Xiang and Li, Yan and Zhang, Yongdong and Wang, Meng},
  booktitle={Proceedings of the 43rd International ACM SIGIR conference on research and development in Information Retrieval},
  pages={639--648},
  year={2020}
}

@article{mao2024matrix,
  title={Matrix factorization recommendation algorithm based on attention interaction},
  author={Mao, Chengzhi and Wu, Zhifeng and Liu, Yingjie and Shi, Zhiwei},
  journal={Symmetry},
  volume={16},
  number={3},
  pages={267},
  year={2024},
  publisher={MDPI}
}

@inproceedings{rashed2022context,
  title={Context and attribute-aware sequential recommendation via cross-attention},
  author={Rashed, Ahmed and Elsayed, Shereen and Schmidt-Thieme, Lars},
  booktitle={Proceedings of the 16th ACM conference on recommender systems},
  pages={71--80},
  year={2022}
}

@inproceedings{lu2021personalized,
  title={Personalized outfit recommendation with learnable anchors},
  author={Lu, Zhi and Hu, Yang and Chen, Yan and Zeng, Bing},
  booktitle={Proceedings of the IEEE/CVF conference on computer vision and pattern recognition},
  pages={12722--12731},
  year={2021}
}

@article{ma2024personalized,
  title={Personalized fashion recommendations for diverse body shapes and local preferences with contrastive multimodal cross-attention network},
  author={Ma, Jianghong and Sun, Huiyue and Yang, Dezhao and Zhang, Haijun},
  journal={ACM Transactions on Intelligent Systems and Technology},
  volume={15},
  year={2024}
}

@inproceedings{zhao2022multi,
  title={Multi-view intent disentangle graph networks for bundle recommendation},
  author={Zhao, Sen and Wei, Wei and Zou, Ding and Mao, Xianling},
  booktitle={Proceedings of the AAAI conference on artificial intelligence},
  volume={36},
  number={4},
  pages={4379--4387},
  year={2022}
}

@inproceedings{zou2022multi,
  title={Multi-level cross-view contrastive learning for knowledge-aware recommender system},
  author={Zou, Ding and Wei, Wei and Mao, Xian-Ling and Wang, Ziyang and Qiu, Minghui and Zhu, Feida and Cao, Xin},
  booktitle={Proceedings of the 45th international ACM SIGIR conference on research and development in information retrieval},
  pages={1358--1368},
  year={2022}
}

@inproceedings{yu2022graph,
  title={Are graph augmentations necessary? simple graph contrastive learning for recommendation},
  author={Yu, Junliang and Yin, Hongzhi and Xia, Xin and Chen, Tong and Cui, Lizhen and Nguyen, Quoc Viet Hung},
  booktitle={Proceedings of the 45th international ACM SIGIR conference on research and development in information retrieval},
  pages={1294--1303},
  year={2022}
}

@inproceedings{rong2024llm,
  title={LLM Enhanced Representation for Cold Start Service Recommendation},
  author={Rong, Dunlei and Yao, Lina and Zheng, Yinting and Yu, Shuang and Xu, Xiaofei and Liu, Mingyi and Wang, Zhongjie},
  booktitle={International Conference on Service-Oriented Computing},
  pages={153--167},
  year={2024},
  organization={Springer}
}

@inproceedings{lee2019melu,
  title={Melu: Meta-learned user preference estimator for cold-start recommendation},
  author={Lee, Hoyeop and Im, Jinbae and Jang, Seongwon and Cho, Hyunsouk and Chung, Sehee},
  booktitle={Proceedings of the 25th ACM SIGKDD international conference on knowledge discovery \& data mining},
  pages={1073--1082},
  year={2019}
}

@article{volkovs2017dropoutnet,
  title={Dropoutnet: Addressing cold start in recommender systems},
  author={Volkovs, Maksims and Yu, Guangwei and Poutanen, Tomi},
  journal={Advances in neural information processing systems},
  volume={30},
  year={2017}
}

@article{sun2024revisiting,
  title={Revisiting bundle recommendation for intent-aware product bundling},
  author={Sun, Zhu and Feng, Kaidong and Yang, Jie and Fang, Hui and Qu, Xinghua and Ong, Yew-Soon and Liu, Wenyuan},
  journal={ACM Transactions on Recommender Systems},
  volume={2},
  number={3},
  pages={1--34},
  year={2024},
  publisher={ACM New York, NY}
}

@article{rendle2012bpr,
  title={BPR: Bayesian personalized ranking from implicit feedback},
  author={Rendle, Steffen and Freudenthaler, Christoph and Gantner, Zeno and Schmidt-Thieme, Lars},
  journal={arXiv preprint arXiv:1205.2618},
  year={2012}
}

@inproceedings{ma2022crosscbr,
  title={CrossCBR: cross-view contrastive learning for bundle recommendation},
  author={Ma, Yunshan and He, Yingzhi and Zhang, An and Wang, Xiang and Chua, Tat-Seng},
  booktitle={Proceedings of the 28th ACM SIGKDD conference on knowledge discovery and data mining},
  pages={1233--1241},
  year={2022}
}

@inproceedings{wu2021self,
  title={Self-supervised graph learning for recommendation},
  author={Wu, Jiancan and Wang, Xiang and Feng, Fuli and He, Xiangnan and Chen, Liang and Lian, Jianxun and Xie, Xing},
  booktitle={Proceedings of the 44th international ACM SIGIR conference on research and development in information retrieval},
  pages={726--735},
  year={2021}
}

@article{cai2023lightgcl,
  title={LightGCL: Simple yet effective graph contrastive learning for recommendation},
  author={Cai, Xuheng and Huang, Chao and Xia, Lianghao and Ren, Xubin},
  journal={arXiv preprint arXiv:2302.08191},
  year={2023}
}

@article{gao2023survey,
  title={A survey of graph neural networks for recommender systems: Challenges, methods, and directions},
  author={Gao, Chen and Zheng, Yu and Li, Nian and Li, Yinfeng and Qin, Yingrong and Piao, Jinghua and Quan, Yuhan and Chang, Jianxin and Jin, Depeng and He, Xiangnan and others},
  journal={ACM Transactions on Recommender Systems},
  volume={1},
  number={1},
  pages={1--51},
  year={2023},
  publisher={ACM New York, NY, USA}
}

\end{document}